# Beta-diversity of Central European forests decreases along an elevational gradient due to the variation in local community assembly processes

Sabatini, Francesco Maria[*]; Jiménez-Alfaro, Borja; Burrascano, Sabina; Lora, Andrea; Chytrý, Milan.

**Addresses**

Sabatini, F.M. - Department of Geography, Humboldt-Universität zu Berlin, Unter den Linden 6, 10099 Berlin, Germany. francesco.maria.sabatini@geo.bu-berlin.de

ORCID ID: 0000-0002-7202-7697

[*] Corresponding author

Jiménez-Alfaro, B. - German Centre for Integrative Biodiversity Research (iDiv) Halle-Jena-Leipzig; Deutscher Platz 5e, 04103 Leipzig, Germany; Martin Luther University Halle Wittenberg, Institute of Biology, Am Kirchtor 1, D-06108 Halle (Saale), Germany. borja.jimenez-alfaro@botanik.uni-halle.de

Burrascano S. - Department of Environmental Biology, Sapienza University of Rome, P.le Aldo Moro 5, 00185 Rome, Italy. sabina.burrascano@uniroma1.it

Lora, A. - Istituto di Cristallografia, C.N.R., via Salaria km 29,300, 00015 Monterotondo, Italy. andrea.lora@ic.cnr.it

Chytrý, M. - Department of Botany and Zoology, Masaryk University, Kotlářská 2, 611 37 Brno, Czech Republic. chytry@sci.muni.cz

**Abstract**

Beta-diversity has been repeatedly shown to decline with increasing elevation, but the causes of this pattern remain unclear, partly because they are confounded by coincident variation in alpha- and gamma-diversity. We used 8,795 forest vegetation-plot records from the Czech National Phytosociological Database to compare the observed patterns of beta diversity to null-model expectations (beta-deviation) controlling for the effects of alpha- and gamma-diversity. We tested whether β-diversity patterns along a 1,200 m elevation gradient exclusively depend on the effect of varying species pool size, or also on the variation of the magnitude of community assembly mechanisms determining the distribution of species across communities (e.g., environmental filtering, dispersal limitation). The null model we used is a novel extension of an existing null-model designed for presence/absence data and was specifically designed to disrupt the effect of community assembly mechanisms, while retaining some key features of observed communities such as average species richness and species abundance distribution. Analyses were replicated in ten subregions with comparable elevation ranges. Beta-diversity declined along the elevation gradient due to a decrease in gamma-diversity, which was steeper than the decrease in alpha-diversity. This pattern persisted after controlling for alpha- and gamma-diversity variation, and the results were robust when different resampling schemes and diversity metrics were used. We conclude that in temperate forests the pattern of decreasing beta-diversity with elevation does not exclusively depend on variation in species pool size, as has been hypothesized, but also on variation in community assembly mechanisms. The results were consistent across resampling schemes and diversity measures, thus supporting the use of vegetation plot databases for understanding patterns of beta-diversity at the regional scale.

**Introduction**

Species diversity is often decomposed in its alpha (α), beta (β) and gamma (γ) components (Whittaker, 1972). These three aspects of diversity are influenced by several factors, including evolutionary history, climatic variation, habitat heterogeneity, dispersal limitation, species interactions and stochastic effects (Condit et al. 2002; Legendre et al. 2009; Chase, 2010; Mori et al. 2015). The interplay of these factors creates striking biodiversity patterns that have long attracted the researchers' attention (Nogues-Bravo et al. 2008). Whereas classical biogeographical studies have been mainly focused on latitudinal and elevational patterns of α- and γ-diversity, interest in patterns of β-diversity is relatively new and the interpretation of these patterns is still subject to scientific debate (Qian and Ricklefs, 2007; Lenoir et al. 2010; Kraft et al. 2011; De Cáceres et al. 2012; Myers et al. 2013, Ulrich et al. 2017).

Besides describing the scaling between α- and γ-diversity, β-diversity (site-to-site variation in species composition) is the result of assembly processes operating at a given geographic scale (Condit et al. 2002; Legendre et al. 2009). Empirical studies focusing on β-diversity are thus fundamental to provide insights into the processes that create and maintain the compositional variation of natural communities (Tuomisto et al. 2003; Chase, 2010; Burrascano et al. 2013; Myers et al. 2013). Although β-diversity has been reported to generally decline with increasing latitude and elevation (Kraft et al. 2011; Mori et al. 2013, Tello et al. 2015), there is no general consensus about the causes of this decline. This uncertainty partly derives from the fact that comparisons of β-diversity among sites or regions are confounded by coincident variation in α- and γ-diversity (Tuomisto, 2010; Chase et al. 2011). The relationship with γ-diversity is particularly important because it is related to evolutionary and historical processes operating at large spatio-temporal scales determining the size of the species pool, i.e. the habitat-specific set of species that can potentially inhabit a site because of suitable ecological conditions (Zobel, 2016).

Using a null-model approach, Kraft et al. (2011) suggested that latitudinal and elevational patterns of β-diversity are related to the concomitant variation in the species pool, concluding that there is no need to invoke differences in the mechanisms of local community assembly to explain these patterns. However, later studies reported a dominant effect of local community assembly mechanisms (Mori et al. 2013; Tello et al. 2015). Moreover, Qian et al. (2013) criticized the null-model used by Kraft et al. (2011), claiming that it retains the species abundance distributions (SADs) observed in real data, disregarding the fact that these are the result of mechanisms of local community assembly and in turn might generate β-diversity gradients (Qian et al. 2013; Mori et al. 2015). This debate is complicated by the fact that the historical and evolutionary regional assembly processes related to the species pools and local assembly processes (e.g., environmental filtering, competition or dispersal limitation), besides being intimately related, are by definition dominant at different spatial scales. Indeed, patterns of species occurrence and abundance depend on the geographic extent of the study system, as well as on grain and sampling schemes (Tello et al. 2015, Tang et al. 2012). Furthermore, the relative importance of different assembly processes may change among biogeographical regions, either due to different evolutionary histories or different ecological conditions (Chase, 2010; Kraft et al. 2011; De Cáceres et al. 2012; Myers et al. 2013; Qian et al. 2013; Myers et al. 2015).

Analyses of β-diversity are often restricted to one or a few elevation transects, which may represent only a small fraction of the regional diversity (Tello et al. 2015). This lack of replication may result in inconsistencies among the outcomes of different studies. Ideally, inference on β-diversity patterns would be improved if well-replicated studies conducted within the same biogeographical context and following a standard methodology were compared (Lessard et al. 2012). Furthermore, conducting studies at a regional scale (~$10^3$-$10^5$ km$^2$) would substantially limit the potential confounding effects of different species pools, evolutionary histories and ecological conditions. Indeed, within the same region, community assembly processes relevant at the level of meta-communities (e.g., dispersal limitation and environmental filtering) have the clearest effect (Leibold et al. 2004; Cottenie, 2005; Jiménez-Alfaro et al. 2015). Despite the numerous advantages of analyzing regional datasets such an approach was until recently hampered by the limited availability of large collections of fine-scale community data. Fortunately, such datasets have been recently made available for plant communities with the development of national and regional vegetation databases (Dengler et al. 2011; Chytrý et al. 2016).

In this study, we used vascular plant species data from 8,795 forest vegetation plots in the Czech Republic to test the consistency of β-diversity patterns across a 1,200 m elevation gradient replicated in 10 adjacent subregions. Specifically, we tested the hypothesis that the variation in the species pool size is the sole factor responsible for variation in β-diversity patterns along the elevation gradient. If this hypothesis is correct, then any relationships between β-diversity and elevation should disappear after controlling for the variation in species pool sizes (Kraft et al. 2011; Lessard et al. 2012; Mori et al. 2015), i.e. the net effect of the main local assembly processes that determine the distribution of species across communities, such as environmental filtering, competition or dispersal limitation, should be constant along elevation gradients. In contrast, if the observed patterns in β-diversity persist after removing the effect of variation in the species pool size, the effect of the assembly processes should change with elevation (Tello et al. 2015). To distinguish between these two possibilities, we tested whether the observed patterns of β-diversity differed from random expectations generated by sampling from a species pool that accounts for the influence of evolutionary and historical processes (Lessard et al. 2012). We used a null-model that was specifically designed to disrupt the effect of those assembly mechanisms that determine the distribution of species across communities, while retaining some key features of observed communities such as average species richness and SADs. We tested the robustness of the observed inference by comparing the results obtained when using the whole dataset vs. three resampled datasets, and when using diversity metrics that give progressively less weight to rare species.

**Methods**

*Study area and vegetation data*

Czech Republic is a land-locked country of central Europe occupying an area of 78,867 km$^2$ and an elevation range between 115 and 1602 m a.s.l. The climate is temperate oceanic to temperate continental (Rivas-Martínez et al. 2004), with continentality increasing from west to east and from mountains to lowlands. Both temperature and precipitation peak in July. Lowlands are warm and dry, with a mean annual temperature of 8–9.5 °C (January mean –2 to 0 °C, July mean 18–20 °C) and annual precipitation of 400–600 mm (Tolasz et al. 2007). The highest areas in the mountains have a mean annual temperature of about 1–2 °C (January mean about –7 to –6 °C, July mean about 8–10 °C) and annual precipitation of 1200–1400 mm.

The Czech National Phytosociological Database contains over 100,000 vegetation plots (relevés), with an estimated density of more than 1,000 plots/1,000 km$^2$ (Chytrý and Rafajová, 2003; Dengler et al. 2011). We restricted the initial set of vegetation plots to include only forest vegetation, thus considering 19,133 plots taken by 354 authors between 1924 and 2012. Only data relative to vascular plants were used. Data contained in the Czech National Phytosociological Database are homogeneous with respect to taxonomic information. The nomenclature follows Ehrendorfer (1973). For controversial taxonomic groups, species or subspecies were transformed to species groups or aggregated species, to avoid confusion arising from the use of different taxonomic concepts from different authors and biases in the analyses (Chytrý and Rafajová, 2003). To further assure the consistency of the dataset, we selected 12,781 vegetation plots (i) sampled after 1980; (ii) with a cover of tree species collectively greater than 30%; (iii) with plot size greater or equal to 100 m$^2$. We replicated the analysis across 10 subregions by aggregating neighbouring phytogeographical districts (Kaplan, 2012), as these are more suitable to explore diversity patterns than administrative units (Abbate et al. 2015). Each subregion encompassed an average surface of 5,155 km$^2$ (range 1,666-10,862) and ranged from lowlands to mountaintops or to the alpine timberline at 1200–1400 m (Fig. 1, Table 1). The final number of vegetation plots included within the ten subregions was 8,795, containing a total of 1,250 species.

*Resampling*

Vegetation-plot databases may produce biased samples of community diversity due to uneven representation of vegetation types or geographical regions, or both (e.g., positive bias towards easily accessible areas or areas of special interest), and non-random location of plots. To consider the potential effect of these biases, we prepared four datasets, one using all the available data, and three using different resampling approaches. In order to control for random variability across realizations, each resampling approach was replicated 100 times and the results averaged (Supplementary material - Fig. S1):

1. No resampling (dataset WHOLE). All the vegetation plots belonging to the 10 subregions were retained. For each subregion, plots were sorted according to their elevation and divided into groups of 20. Groups encompassed uneven elevational ranges, especially at the tails of the distribution of elevations. We set a threshold of 200 m as the maximum elevational span accepted for each group, and discarded the plots located at the lower and upper tail of the distribution of elevations until this threshold was reached.

2. Elevational resampling (ALTBIN). Each vegetation plot was assigned to a 100 m wide elevational belt, and the number of plots in each subregion*elevational belt was calculated. For each subregion, only elevational belts having more than 40 plots were considered. If plots did not encompass the whole elevational range of a belt in a subregion (e.g., when the lowest or highest plot of a subregion occurred at an elevation respectively higher than 130 m or lower than 170 m within a 100-200 m elevational belt), that belt was excluded. In each replication, we randomly resampled a group of 20 plots for each valid belt*subregion combination.

3. Geographic resampling (GRID). Plots were resampled within a 1 x 1 km square grid. For each subregion, we randomly drew one plot for each cell of the grid containing plots. Plots were then sorted according to their elevation, and divided into groups of 20 as in the dataset WHOLE. As above, we set a threshold of 200 m as the maximum elevational span accepted for each group.

4. Elevational and geographic resampling (BINGRID). A combination of resampling schemes 2 and 3, in which resampling was constrained using 100 m elevational belts, with an additional constraint that no more than one plot per elevational belt could be drawn for each 1 km$^2$ cell. Plots were then divided into groups of 20 as in the dataset ALTBIN.

The number of groups ranged from 52 (dataset BINGRID) to 439 (dataset WHOLE), and the groups were heterogeneously distributed across the subregions*resampling combinations, ranging from 0 to 76 groups per subregion (Table S1).

*Diversity calculations*

In this work, we quantified β-diversity (*sensu latu*) using two complementary approaches. First we calculated β-diversity as the effective number of compositionally distinct sampling units (i.e. vegetation plots) in a group. This equals true β-diversity as defined by Tuomisto (2010) based on the foundations laid by Hill (Hill, 1973; Jost, 2007). For each group of plots, we calculated species diversity as follows:

$$^qD = \frac{1}{\sqrt[q-1]{\sum_{i=1}^{s} p_i p_i^{q-1}}} \quad (1)$$

where $p_i$ is the proportional abundance of species *i*, *S* is the total number of species, and *q* is the order of the diversity. We calculated species diversity using *q*=[0,1,2]. When *q*=0, species abundances are cancelled out from the equation, so $^0D$ obtains the same numeric value as species richness. For increasing *q*, abundant species are given progressively more weight than implied by their proportional abundances. Using different biodiversity metrics that give progressively less weight to rare species allows us to consider the potential effect of undersampling, which depends on sampling effort (e.g., number of observers and observation time) and completeness (i.e. the ratio between observed and actual richness) and is a common problem in vegetation databases. This approach assumes that rare species are more prone to undersampling than common species (Cardoso et al. 2009; Beck et al. 2013). For each group of plots, we followed Tuomisto (2010) and performed a multiplicative partitioning of the total species diversity observed: $^qD_\gamma = {^qD_\alpha} \times {^qD_\beta}$. $^qD_\gamma$ represents the overall diversity of the group of plots, $^qD_\alpha$ the average α-diversity, and $^qD_\beta$ their overall β-diversity, for a given order of diversity *q*. We decomposed diversity using the R script provided in Sabatini et al. (2014).

As a second approach, we followed Legendre and De Cáceres (2013) and summarized compositional heterogeneity as the variation of plot-to-plot dissimilarity matrices calculated for different levels of diversity *q* ($^q\Delta$), as:

$$Var(^q\Delta) = \frac{1}{n} \sum_{h=1}^{n-1} \sum_{i=h+1}^{n} {^q\Delta_{hi}^2}$$

where *h* and *i* represent the position index in the subdiagonal of $^q\Delta$ and n is the number of plots (not the number of distances). Although correlated to β-diversity ($^qD_\beta$), $Var(^q\Delta)$ is not trivially determined by matrix fill, i.e. the proportion of occupied cells in a species × sites matrix, which can be shown it is the inverse of $^0D_\beta$ (Ulrich et al. 2017), but depends also on the distribution of species across sites.

*Null model approach for calculating β-deviation*

We developed a null-model to cancel the dependency of β-diversity on α- and γ-richness, by removing the effects of those assembly mechanisms that determine the distribution of species across communities, but keeping local species richness and SADs as they are in the study system. For each group of plots we generated a matrix of species composition, $Y_{obs}$, which contained the cover values of the species observed in each plot. For each $Y_{obs}$ we generated 999 null data tables ($Y_{perm.i}$) having the same dimensions as $Y_{obs}$. This means in practices, that the species pool was defined as all the species that are found within a given subregion and elevation band, therefore accounting for the influence of regional scale evolutionary and historical processes (Lessard et al. 2012). Species occurrences were permuted among plots in the group using the proportional-proportional (PP) algorithm described in Ulrich and Gotelli (2012), separately for each vegetation layer (i.e. overstorey vs. herb-layer). The PP algorithm creates presence-absence permuted data table in which rows and columns vary randomly, but the average row and column totals match those of the observed community data. This 4-step algorithm first assigns matrix row and column totals from a binomial distribution centered around the observed total for each species and site, adjusting the marginal totals to avoid small differences in row and column totals, if necessary. Matrix cell occurrences are then placed step by step following a proportional-proportional null model, and multiple entries are reduced by the sum-of-squares algorithm (SSR) of Miklós and Podani (2004). In those relatively rare cases when the PP algorithm defines an impossible matrix state, and the SSR algorithm runs into a dead end, these irreducible multiple entries are placed into empty cells proportional to the predefined marginal totals (Ulrich and Gotelli 2012).

After permuting the species occurrences, the cover values of each species occurring in each site of a permuted matrix were sampled with replacement from the vector of observed abundances of the specific species within the group of plots. Permuted cover values of each site (i.e. row vectors) were finally scaled to match the overall sums of species cover values of observed community (i.e. row totals). This null model was designed to meet four constraints: (1) to retain the average number of species of each plot (i.e. average 'α-richness' or $^0D_\alpha$) and the overall number of species in the species pool (i.e. 'γ-richness' or $^0D_\gamma$) of each group of plots; (2) to keep the same proportion of total species richness (and cover) across the tree and understorey layers within each plot; (3) to hold constant the sum of the cover values observed at each plot (as a rough proxy for its overall productivity); and (4) to generate a permuted species abundance distribution (SAD) being centered on, but not identical to, the SAD of observed data (Fig. S2). Here we defined SAD as the vector of abundances (i.e. cover values) of each species in a group of plots. Since no consensus has been achieved on whether SAD should be retained in null models when biodiversity patterns are explored (Mori et al. 2015; Tello et al. 2015), we also ran the analysis (limited to the WHOLE dataset) using a slightly modified null model disrupting the SAD. In this case, the cover values of each species occurring in a site of a permuted matrix were sampled with replacement from the vector of cover values occurring in the corresponding site of the observed matrix.

For each $Y_{obs}$ and each of the 999 permutations of $Y_{perm.i}$, we built a dissimilarity matrix for each order of diversity $q$ (respectively $^q\Delta_{obs}$ and $^q\Delta_{exp.i}$). Dissimilarities between pairs of sites were calculated using the complement of the generalization of the Jaccard index for quantitative data (Jost, 2007; Tuomisto, 2010). This index of species turnover is a monotonic transformation of true β-diversity between two sites as shown below (eq. 2):

$$^qC_\beta = \frac{2}{^qD_\beta} - 1 \qquad (2)$$

We summarized the compositional heterogeneity of each observed and permuted group of vegetation plots calculating the variation of $^q\Delta_{obs}$ and $^q\Delta_{exp}$ as described above (Legendre and De Cáceres 2013). To estimate the influence of the species pool, we calculated the deviation of the observed compositional heterogeneity - $Var(^q\Delta_{obs})$ - from its expected value under the null model as a standardized effect size, i.e. $β_{dev}$ (hereafter β-deviation) was defined as $Var(^q\Delta_{obs})$ minus the average $Var(^q\Delta_{exp})$, divided by the standard deviation of $Var(^q\Delta_{exp})$ across the 999 permutations (Kraft et al. 2011; Mori et al. 2013; Tello et al. 2015). The advantage of using $Var(^q\Delta)$ is that, when matrix fill is held constant as in the case of our null model, $Var(^q\Delta)$ can still vary and its deviation from expected accounts for the amount of compositional heterogeneity not depending on matrix fill.

*Modelling the relationship between β-diversity and β-deviation and elevation*

In total we considered 12 datasets, i.e. the combination between four resampling schemes and three diversity metrics. For the WHOLE dataset, as well as for each of the 100 replications of each resampled dataset, we modelled the relationship between, respectively, $Var(^q\Delta_{obs})$, $Var(^q\Delta_{exp})$, and $β_{dev}$ and the mean elevation of the groups of plots. For $Var(^q\Delta_{obs})$ and $Var(^q\Delta_{exp})$ we used generalized linear mixed models (GLMMs) with a logarithmic link function and assuming a gamma distribution of errors. The response of $β_{dev}$ to elevation, instead, was modelled assuming a normal distribution of errors and an identity link function, i.e. using linear mixed-effect models. Subregions were treated as a random effect (random intercept). We also tested whether modelling mean elevation as a random effect (i.e. using either a random intercept and slope model) improved the fit of the model to the data. Since the uneven distribution of vegetation plots across the study region could have led to subregions including groups of plots being closer or more environmentally similar than in other subregions, we also included some covariates related to climatic and topographical variability, and to the geographical spread of plots within each group. Fixed effects included four sets of descriptors:

set 1) Mean elevation - i.e. the average elevation of the plots included in each group; set 2) Topographical variability - proxied by the standard deviation of the above-canopy annual potential solar irradiation (calculated as a function of latitude, aspect and slope) calculated across the plots of each group (McCune and Keon, 2002); set 3) climatic variability – the standard deviation of mean annual temperature (3a) and the standard deviation of total annual precipitation (3b) of the plots in each group; set 4) geographical spread of the plots within each group – calculated either as (4a) the average between-plot geographical distances calculated for the 20 plots in each group; or proxied by the elevational range (4b) between the lowest and highest plot of a group; or by the geographical extent encompassed by an elevational belt (4c), i.e. the surface area included in the elevational range of a given subregion, calculated on a 20-m resolution digital terrain model; or by the actual forest area (4d) included in the elevational range of a given subregion. Each explanatory variable was standardized to zero mean and unit standard deviation. Variables 3a and 4c were highly correlated to variables 3b and 4d, respectively (Spearman's $\rho$=0.65 and $\rho$=0.80, respectively. p<0.001 for both), and were therefore excluded from subsequent analysis.

For each dataset, we first fitted two global models including all the fixed effects and either a random intercept or a random intercept and slope. We then used the 'dredge' function in the 'MuMIn' (version 1.15.6) R package to fit all the possible combinations of models nested in the global models. Model selection was performed using an Information-Theory approach (Burnham and Anderson, 2002), based on Akaike Information Criterion values corrected for small sample size (AICc). For the WHOLE dataset, the 95% best fitting model set (i.e. the models encompassing 95% of the Akaike weights) was used for multimodel inference and model averaging of parameter estimates. For the resampled datasets we used a slightly different procedure. For each replication we fitted and ranked the global model and the submodels as above but only the top ranking model was retained. Parameter estimates were averaged across the 100 replications and the confidence intervals were calculated as the average ± 1.96 · SE, with the standard error also calculated across the 100 replications. Regression parameters were averaged using the zero-method ('shrinkage towards zero'); a parameter estimate of zero was substituted into those models where a given variable was absent (Burnham and Anderson, 2002; Grueber et al. 2011).

**Results**

*Patterns of β-diversity across resampling schemes and diversity metrics*

Diversity decreased with mean elevation, showing a consistent trend across different orders of diversity *q* (Fig. 2 for the WHOLE dataset) and resampling schemes (Table S2 and Figs. S3, S4, S5).

Because γ-diversity had a steeper decline with elevation than α-diversity (Fig. 2, top row), β-diversity ($^qD_β$) declined with elevation as well (Fig. 2, bottom row).

After averaging across the top-performing GLMMs, the data supported the hypothesis of a declining pattern of observed compositional heterogeneity - $Var(^qΔ_{obs})$ - with increasing elevation across all datasets (Fig. 3, top row for the WHOLE dataset). Other covariates were also comprised in the final, averaged models for the different resampling schemes (Fig. 4; Tables S2, S3). For the WHOLE dataset, observed compositional heterogeneity increased with increasing average between-plot geographical distance and when groups of plots spanned across a wider elevational range (Fig. 4, top row). For the ALTBIN dataset, the top performing models included, besides mean elevation, also forest area, i.e. the extent of forest occurring within the elevational range of a given group of plots in a given subregion. Nevertheless, the confidence intervals of the latter regression coefficient were significantly different from zero only for $q$=0. With minor differences, results for the datasets GRID and BINGRID were qualitatively similar to those of the datasets WHOLE and ALTBIN, respectively and therefore these results are reported in the supplementary material (Tables S2-S5, Fig. S6-S7).

*Observed versus expected β-diversity*

Similarly to observed compositional heterogeneity, expected compositional heterogeneity - $Var(^qΔ_{exp})$ – also decreased with increasing elevation (Fig. 3). The best-fitting models describing the relationship between observed and expected compositional heterogeneity, and the explanatory variables were very similar for each $q$ (results not shown for $Var(^qΔ_{exp})$ ). However, observed compositional heterogeneity decreased more steeply than expected, and as a result, β-deviation also decreased with increasing elevation (Fig. 3, bottom row).

In the WHOLE dataset, β-deviation decreased with increasing altitude for all orders of diversity $q$. β-deviation also increased with increasing average between-plot geographical distance. It also decreased with increasing topographical variability (proxied by the standard deviation of the above-canopy annual potential solar irradiation, Fig. 5, Tab. S5), when $q$=[1,2], and increasing variability in precipitation (when $q$=0). None of these covariates, however, had a regression coefficient different from zero in the ALTBIN dataset, for which the top performing models only included mean elevation as fixed effects (Table S4). Interestingly, across all the datasets, the magnitude of the relationship between β-deviation and elevation progressively decreased when increasing the order of diversity $q$, i.e. when calculating diversity indices assigning less weight to rare, as compared to abundant species.

Both in the case of observed compositional heterogeneity - $Var(^qΔ_{obs})$ - and β-deviation, random intercept and slope models were better supported than random intercept models, especially for the WHOLE dataset, indicating that the relationship between β-deviation and elevation was not constant across the 10 subregions considered (Table S4). Finally, the results obtained when using a null model that disrupted SAD were qualitatively similar to those obtained when retaining SAD (Fig. S8).

**Discussion**

*The influence of assembly processes on elevational gradients in plant β-diversity*

We found that both β-diversity and observed compositional heterogeneity - $Var(^{q}\Delta_{obs})$ - declined with increasing elevation. As expected, observed compositional heterogeneity also increased with geographical spread of vegetation plots and with their elevational range, indicating that the farther apart, or the more environmentally different the plots in a given group are, the higher is their compositional dissimilarity. A declining pattern of β-diversity with increasing elevation has been observed in several tree communities from different biomes (Kraft et al. 2011; Mori et al. 2013; Tello et al. 2015), and here we showed that a similar pattern occurs when extending the analysis from the tree-layer alone, to whole vascular plant assemblages. Although this result is in agreement with the notion that β-diversity is higher in areas with higher productivity (Chase 2010), exceptions have also been reported, suggesting that the mechanisms causing β-diversity patterns may differ across scales (Tang et al. 2012) or geographical regions (De Cáceres et al. 2012, Qian et al. 2013).

Interestingly, the elevational gradients in compositional heterogeneity persisted after controlling for the size of the species pools, as indicated by the systematic variation in β-deviation with elevation. This result implies that the overall magnitude of those assembly processes that determine the distribution of species across communities, such as environmental filtering, competition or dispersal limitation, varied along the ~1,200 m elevational gradient existing in the study area, and influenced the elevational pattern of β-diversity. Although we found that β-deviation consistently decreased with elevation, our data showed that the slope of this gradient varied across subregions, supporting the conclusion of Qian et al. (2013) that analyses of biodiversity patterns should be conducted separately for different geographic areas.

The deviation from expected β-diversity is usually attributed to multiple ecological processes, including either spatial (e.g., dispersal limitation), environmental (e.g., species sorting), or stochastic (e.g., ecological drift or priority effect) (Mori et al. 2015; Tello et al. 2015) effects. Observed communities were more diverse than expected at low elevation, and less diverse than expected at high elevation. This may imply that the assembly processes determining species turnover in forest communities are stronger at lower than at higher elevations. At low elevation, for instance, communities may be more strongly shaped by environmental filtering and dispersal limitation than at high elevation, because of higher habitat heterogeneity or wider geographical area. The higher than expected β-diversity may also be the results of other mechanisms being stronger in productive, species rich communities at lower elevation, including competition or priority effect (Chase 2010), or depend on a more widespread and heterogeneous effect of human disturbance and forest management. Alternatively, the higher than expected β-diversity at low elevation may depend on the species ecological ranges of plant assemblages occurring at low vs. high elevations. The classical Rapoport-Rescue Hypothesis, for instance, suggests that the greater diversity that occurs at low latitudes (e.g., in the tropics) depends on the fact that species have narrower ecological ranges at lower latitudes (Stevens 1989; Willig et al. 2003; Qian and Ricklefs, 2007). Whether a similar argument could be advanced when considering lowland vs. mountain communities is, however, controversial. If high elevation species had, on average, wider ecological ranges than low elevation species, communities should be compositionally more homogeneous at high elevation than in the lowlands just as a result of chance. The fact that when less weight is assigned to rare species (i.e. for high q), β-deviation decreased less steeply with elevation, may provide some indirect support to this hypothesis, given that specialist species have usually more restricted geographical and ecological ranges than common species. Although further research is undoubtedly needed, we believe that vegetation-plot databases may represent powerful instruments for testing such a hypothesis.

*The role of null models for exploring β-diversity patterns*

The null-model approach we developed is a novel extension of the proportional-proportional algorithm (Ulrich and Gotelli, 2012), here used for the first time in the context of exploring β-diversity patterns, not only of woody species but of whole vascular plant assemblages. Plant cover (rather than number of individuals) is the most commonly used abundance measure for non-tree plants in vegetation surveys, and consequently, it is also common in vegetation-plot databases. Developing a null model for these data was needed given the increasing potential of the use of these databases for the exploration of diversity gradients (Mori et al. 2015). Furthermore, our novel approach overcomes concerns regarding the use of null models in the context of exploring biodiversity patterns. Traditional null modeling approaches, indeed, are prone to the so-called "Narcissus effect", i.e., an artificially high similarity between the observed and expected species distribution that can potentially lead to inflated rejection rates for focal patterns (Lessard et al. 2012; Ulrich et al. 2017). The advantage of the PP algorithm is that it relaxes the usual constraint of many fixed-fixed null models (e.g., the 'trial-swap' algorithm, Miklós and Podani, 2004) that create permuted matrices having row and column totals identical to those of the original matrix. The PP model, instead, retains the desired levels of average local species richness (mean α-richness) and the desired species pool sizes, but without inflating the similarity between observed and permuted species distribution. By coupling this approach with a metric of compositional heterogeneity not trivially related to matrix fill, we were able to model how compositional heterogeneity varied along an elevational gradient, for a given species pool size and average species richness.

Null models should be constructed to deliberately exclude the mechanisms that generate the pattern under investigation (Gotelli 2001; Qian et al. 2013). Similarly to Kraft et al. (2011), our randomization algorithm was specifically designed for disrupting the mechanisms causing species co-occurrence, while controlling for SAD, and for plot- and species pool richness. Whether SAD is driven by the same mechanisms of local community assembly that generate β-diversity gradients and whether it can exert a control on the degree of deviation from the expected β-diversity is a controversial issue (Qian et al. 2013; Morti et al., 2015; Xu et al. 2015). Here, we generated randomized communities that had a similar, but not identical, pattern of the observed SAD. This is a desirable property of a null model, since creating randomized communities where rare species are unrealistically dominant, and dominant species rare, may not return ecologically meaningful results (Mori et al. 2015). Although, we acknowledge that the desirability of retaining SAD in null models requires more research, our results did not depend heavily on the effect of the SAD. Indeed, our findings were robust both when considering different orders of diversity $q$ (indeed, when $q$=0, only species occurrences are considered and SAD is simply disregarded), and when using a null model that disrupts SAD. Altogether, this suggests that SAD plays only a minor role in β-deviation patterns in central European forests.

*Resampling vegetation databases to infer regional patterns of β-diversity*

Studies of β-diversity patterns have been frequently limited by the lack of a sufficient number of replicates of elevation transects (Tello et al. 2015). Vegetation-plot databases are becoming more available and accessible to ecologists, and in the coming years they will help overcome data limitations and allow studies at unprecedented spatial scales (Dengler et al. 2011; Chytrý et al. 2016). By using different resampling schemes, we took into account possible sources of bias that may be associated with such databases, such as uneven geographical distribution of sampling units and preferential sampling. In general, the results concerning the elevation gradients in β-diversity and β-deviation were qualitatively similar between the WHOLE and the resampled datasets, indicating that the database we used is sufficiently robust to these biases. This indicates that the potential sources of bias in vegetation-plot databases have only a minor confounding effect on the outcome of empirical studies aimed at exploring geographical biodiversity patterns, as long as fairly strong ecological gradients are investigated. Interestingly, performing analyses on data resampled without regards to elevation (i.e. the GRID dataset) produced results more similar to those from the unresampled, WHOLE dataset than when performing the same analysis on datasets that were explicitly resampled to account for the vertical distribution of vegetation plots (i.e., datasets ALTBIN and BINGRID). Furthermore, when modeling elevational patterns of β-diversity and β-deviation, the latter resampling schemes returned regression coefficients having a confidence interval overlapping zero for most of the covariates. This indicates that these resampling schemes successfully controlled for the bias deriving from the uneven distribution of vegetation plots in the geographic or environmental space, and confirms the need of adopting schemes specifically designed in function of the objective of the analysis (Knollová et al. 2005).

**Conclusions**

We analyzed a comprehensive vegetation-plot database of forest plant communities from Central Europe to show that observed β-diversity declined with elevation due to a decrease in γ-diversity, which was steeper than the decrease in α-diversity. This elevation gradient persisted even after using a null model that controlled for the confounding variation in species pool size and that disrupted the mechanisms causing species co-occurrence at the local scale. The deviation of observed β-diversity from its expected values under the null model suggests that the magnitude of different local community assembly mechanisms changes along the elevation gradient considered, and that the gradient of β-diversity was not caused exclusively by the decline in species pool size with increasing elevation. This means that the relative importance of those assembly processes that lead to species turnover in forest communities are stronger at lower rather than at higher elevations, possibly as a result of a stronger effect of environmental filtering, dispersal limitation or priority effects at lower elevation.

**Acknowledgements**

The project was funded by two grants to F.M. Sabatini from Sapienza University of Rome (Grant no. C26N12ESSE, C26N14JZMA) and a grant to M. Chytrý from the Czech Science Foundation (Grant no. 14-36079G). We would like to thank Prof. Carlo Blasi for scientific advice and support throughout the research, Giovanni Strona and Nicholas Gotelli for sharing their code for the PP null model, Matthias Baumann for help with recoding the PP null model into R. Finally, we are grateful towards the subject editor and the two anonymous reviewers for their constructive criticisms that helped us to strongly improve this paper.

**Supplementary material**

Appendix 1 – R script for calculating the proportional-proportional null model for presence\absence data.
Table S1 – Diversity estimations across subregions and elevational belts.

Tuomisto, H. 2010. A diversity of beta diversities: straightening up a concept gone awry. Part 2. Quantifying beta diversity and related phenomena. - Ecography 33: 23-45.

Ulrich, W. et al. 2017. The tangled link between β-and γ-diversity: a Narcissus effect weakens statistical inferences in null model analyses of diversity patterns. - Glob. Ecol. Biogeogr. 26: 1-5.

Ulrich, W. and Gotelli, N. J. 2012. A null model algorithm for presence–absence matrices based on proportional resampling. - Ecol. Model. 244: 20-27.

Whittaker, R. H. 1972. Evolution and measurement of species diversity. - Taxon 21: 213-251.

Willig, M. R. et al. 2003. Latitudinal Gradients of Biodiversity: Pattern, Process, Scale, and Synthesis. - Annu. Rev. Ecol. Evol. Syst. 34: 273–309.

Xu, W. et al. 2015. Latitudinal differences in species abundance distributions, rather than spatial aggregation, explain beta-diversity along latitudinal gradients. - Glob. Ecol. Biogeogr. 24: 1170-1180.

Zobel, M. 2016. The species pool concept as a framework for studying patterns of plant diversity. - J. Veg. Sci. 27: 8-18.
19

**Tables**

**Table 1** – Brief description of the subregions considered in this study. T = temperature.

| Subregion Name | Area (km²) | Elevation (m) | | Mean T Jan (°C) | | Mean T Jul (°C) | | Mean Annual T (°C) | | Annual Precipitation (mm) | | No. of Plots |
|---|---|---|---|---|---|---|---|---|---|---|---|---|
| | | min | max | min | max | min | max | min | max | min | max | |
| 1. Krušné hory (Ore Mountains) | 4088 | 154 | 1213 | -4.5 | -0.9 | 12.2 | 18.8 | 3.8 | 8.9 | 450 | 1123 | 430 |
| 2. Slavkovský les | 3161 | 355 | 847 | -4 | -1.8 | 14 | 17.5 | 5 | 7.9 | 490 | 802 | 96 |
| 3. Šumava (Bohemian Forest) | 5788 | 372 | 1366 | -5.5 | -1.9 | 11.8 | 18 | 2.7 | 8.3 | 534 | 1464 | 1267 |
| 4. Brdy | 3544 | 187 | 860 | -3.6 | -0.2 | 14.7 | 19.2 | 5.6 | 9.8 | 489 | 813 | 1482 |
| 5. Krkonoše (Giant Mountains) | 5012 | 202 | 1196 | -5.8 | -1.5 | 10.8 | 18.5 | 2.4 | 8.9 | 576 | 1445 | 761 |
| 6. Orlické hory (Eagle Mountains) | 1666 | 239 | 1078 | -5.4 | -1.9 | 12.1 | 18.2 | 3.4 | 8.6 | 601 | 1350 | 276 |
| 7. Žďárské vrchy | 4978 | 184 | 792 | -4.1 | -1.1 | 15.1 | 18.8 | 6 | 9.1 | 546 | 794 | 217 |
| 8. Jihlavské vrchy | 10862 | 150 | 724 | -3.8 | -1.4 | 15.8 | 19.5 | 6 | 9.6 | 467 | 751 | 1530 |
| 9. Hrubý Jeseník | 7664 | 204 | 1349 | -6.4 | -1 | 10.7 | 18.7 | 1.9 | 9.1 | 552 | 1162 | 1153 |
| 10. Moravskoslezské Beskydy (Moravian-Silesian Beskids) | 4788 | 158 | 1285 | -5.2 | -1.3 | 12.6 | 19.4 | 3.9 | 9.6 | 499 | 1328 | 1483 |

**Figure Legends**

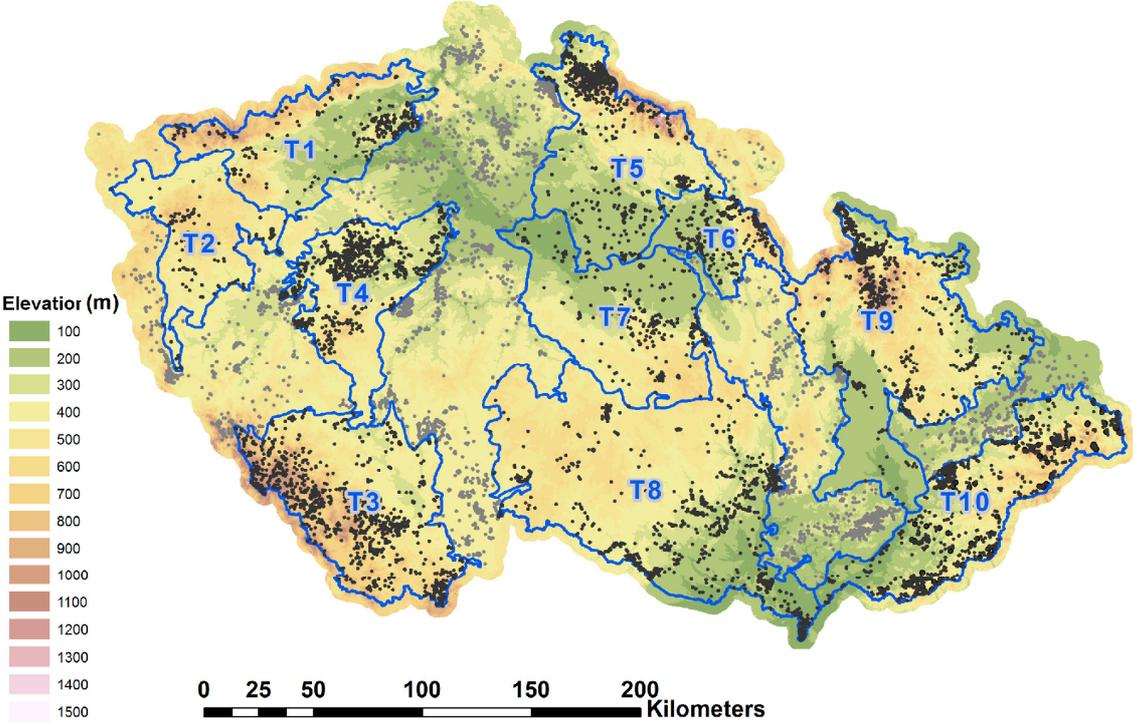

**Fig. 1** – Subregions considered and distribution of the forest plots included in the analysis (black points) and those outside the subregions (white).

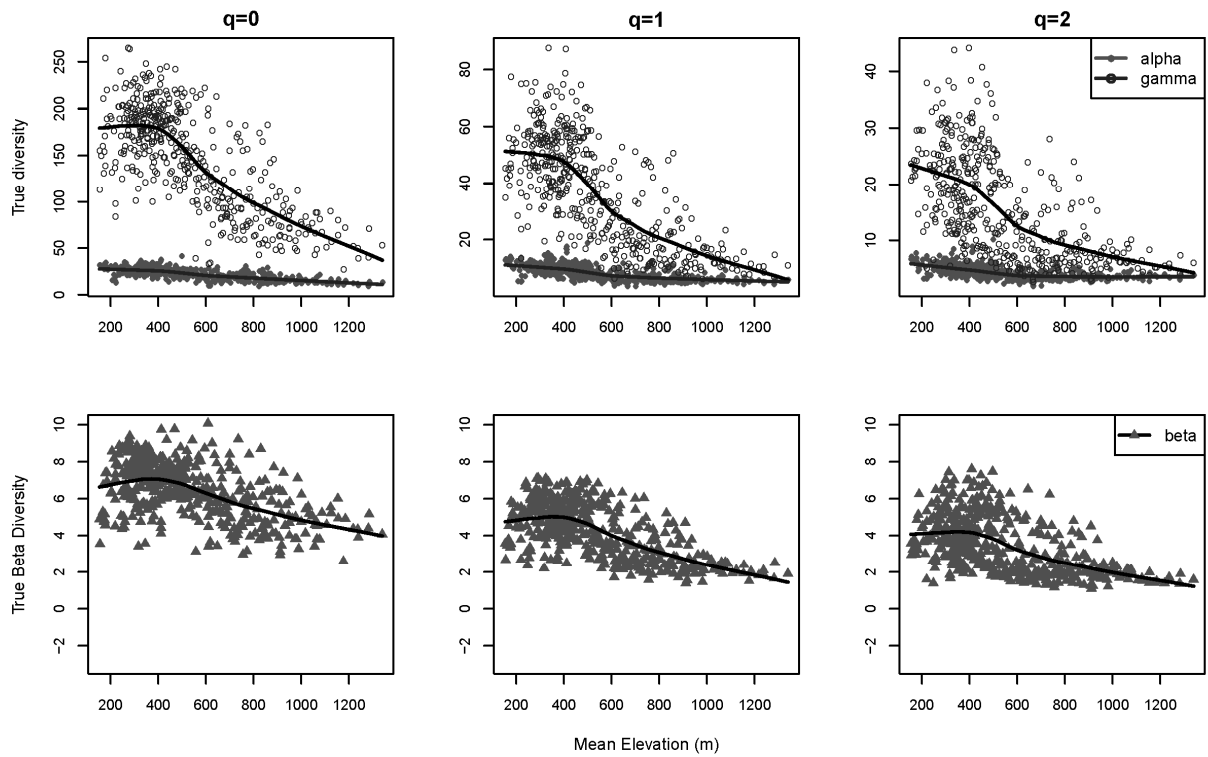

**Fig. 2** – Elevational patterns of α-, γ- (top row) and β-diversity (bottom row) across all the subregions for the WHOLE dataset. Each point represents a group of 20 plots from the same subregion and elevational belt. The trend lines were fitted using a LOWESS smoother.

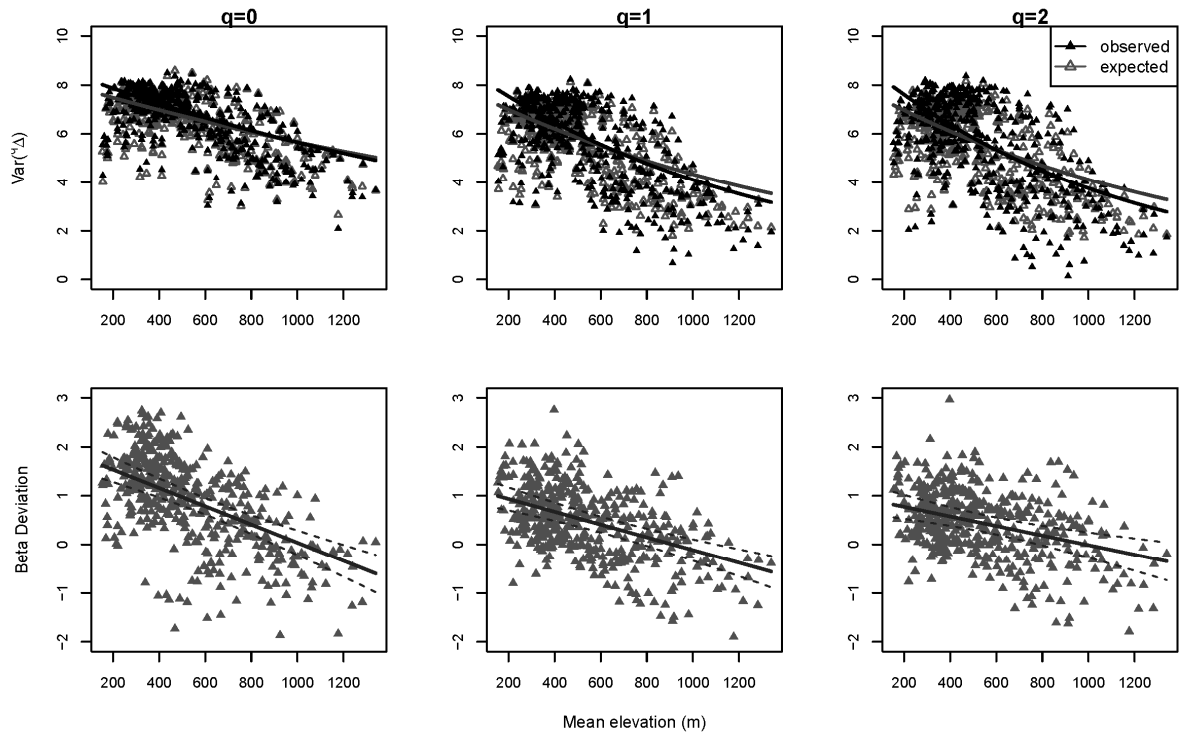

**Fig. 3** – Patterns of observed (top row, black triangles) and expected compositional heterogeneity (top row, grey triangles), and β-deviation (bottom row), along the elevational gradient when considering the WHOLE dataset across 10 elevational subregions in the Czech Republic. Columns represent different orders of diversity *q*. Each symbol represents a group of 20 plots. Regression lines were model-averaged across the 95% best fitting GLMMs.

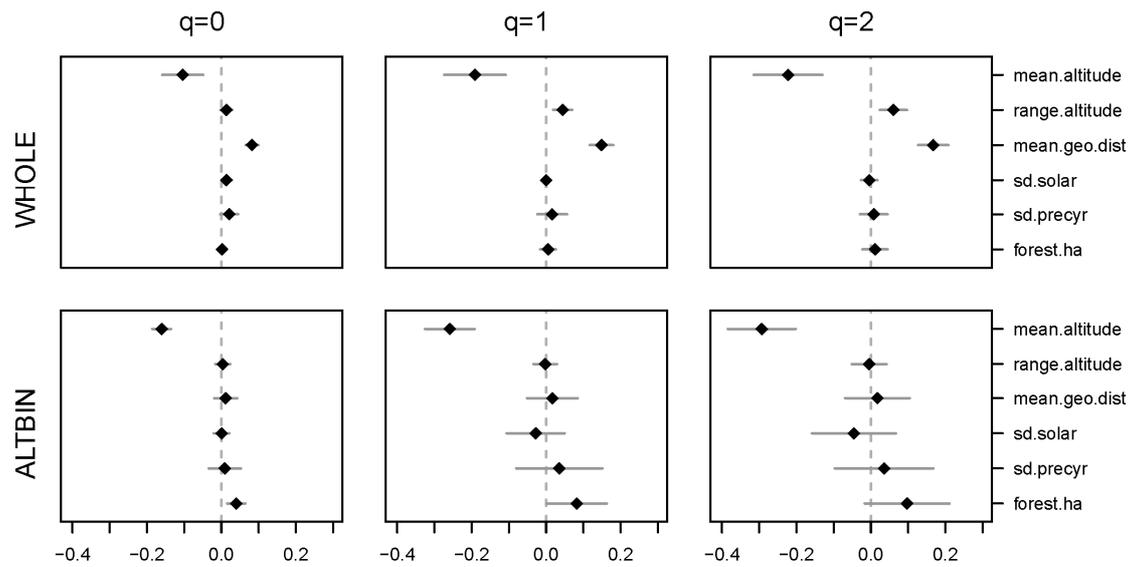

**Fig. 4** – Regression coefficients and confidence intervals of the GLMM (Poisson error distribution, log link function) of the response of observed compositional heterogeneity as a function of elevation, the geographical distribution and the topographical and climatic variability within the groups of forest vegetation plots. Confidence intervals were calculated as the average ± 1.96 SE (with SE estimated across 100 replications for the resampled datasets). Rows represent different resampling schemes.

Explanatory variables – *mean.elevation*: Mean elevation - i.e. the average elevation of the plots included in each group of plots; *range.elevation*: elevational range encompassed by a group of plots; *mean.geo.dist*: geographical spread of the plots within each group of plots calculated as the average between-plot geographical distance; *sd.solar*: standard deviation of the above-canopy annual potential solar irradiation; *sd.prec.yr*: standard deviation of total annual precipitation; *forest.ha*: actual forest area included in the elevational range of a given group of plots.

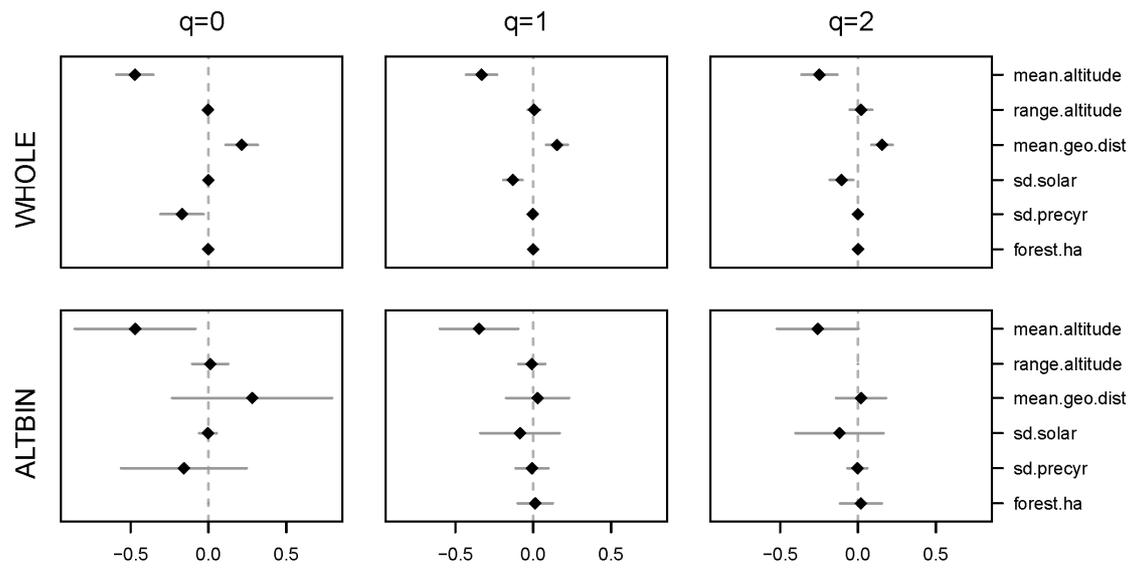

**Fig. 5** – Regression coefficients and confidence intervals of the linear mixed effect model (normal error distribution, identity link function) of the response of the β-deviation as a function of elevation, the geographical distribution and the topographical and climatic variability within the groups of forest vegetation plots. See Fig. 4 for detailed explanations and abbreviations.